\documentclass[pre,aps,reprint,twocolumn,footinbib,longbibliography]{revtex4-1}
\usepackage{graphicx}
\usepackage{amssymb}
\usepackage{amsmath}
\usepackage{times}
\usepackage{bm}

\begin{document}

\title{Nonuniversal large-size asymptotics of the Lyapunov exponent in turbulent globally coupled maps}
  \author{David Velasco}
  \affiliation{Instituto de F\'{\i}sica de Cantabria (IFCA), CSIC-Universidad de
  Cantabria, 39005 Santander, Spain}
  \author{Juan M. L\'opez}
  \affiliation{Instituto de F\'{\i}sica de Cantabria (IFCA), CSIC-Universidad de
  Cantabria, 39005 Santander, Spain}
  \author{Diego Paz\'o}
 \affiliation{Instituto de F\'{\i}sica de Cantabria (IFCA), CSIC-Universidad de
 Cantabria, 39005 Santander, Spain}

\date{\today}

\begin{abstract}
Globally coupled maps (GCMs) are prototypical examples of high-dimensional 
dynamical systems. Interestingly, GCMs formed by an ensemble of weakly coupled identical chaotic units generically exhibit a hyperchaotic `turbulent' state. 
A decade ago, Takeuchi {\em et al.} [Phys.~Rev.~Lett.~{\bf 107}, 124101 (2011)]
theorized that in turbulent GCMs the largest Lyapunov exponent (LE), $\lambda(N)$,
depends logarithmically
on the system size $N$: $\lambda_\infty-\lambda(N)\simeq c/\ln N$.
We revisit the problem and analyze, by means of analytical and numerical 
techniques, turbulent GCMs with positive multipliers to show that there is
a remarkable lack of universality, in conflict with the previous prediction.
In fact, we find a power-law scaling
$\lambda_\infty-\lambda(N)\simeq c/N^{\gamma}$, where $\gamma$ is a 
parameter-dependent exponent in the range $0<\gamma\le1$.
However, for strongly dissimilar multipliers, the LE varies with $N$
in a slower fashion,
which is here numerically explored.
Although our analysis is 
only valid for GCMs with positive 
multipliers, it suggests that a universal 
convergence law for the LE cannot be taken for granted in 
general GCMs.
\end{abstract}

\maketitle

\section{Introduction}

Scaling laws pervade physics. In particular, in the field of chaos theory,
universal routes to low-dimensional chaos with specific 
scaling properties were 
already discovered long time ago~\cite{berge}.
In contrast, high-dimensional chaos--- observed in systems with many ``active''
degrees of freedom--- remains only partly understood~\cite{Cencini}, and scaling laws are often based on more or less heuristic arguments.
Regarding discrete time systems, certain scaling laws have been found 
for coupled-map lattices~\cite{bohr,pik98,cecconi99} and globally coupled
maps (GCMs)~\cite{shibata99,takeuchi11}. 

Concerning GCMs, a rich repertoire of phenomena are known~\cite{kaneko_phd90,kaneko15},  including multistability, clustering, 
chimera or turbulence. Here, we focus on the turbulent regime in GCMs
found at weak coupling. 
The study of turbulent GCMs extends over several decades. 
An early striking discovery 
was the nonstationarity of the mean field in the infinite size limit~\cite{kaneko90,pk94,nakagawa98,kaneko15}. 
Subsequently, several papers characterized the collective properties of chaos 
in turbulent GCMs~\cite{losson98,shibata98,cencini99,takeuchi13}.
Finally, 
Takeuchi et al.~\cite{takeuchi11} uncovered the
delicate arrangement of the Lyapunov exponents underlying turbulent GCMs: 
The Lyapunov spectrum is apparently extensive, but
``subextensive bands'' persist for arbitrarily large system sizes at both ends of the Lyapunov spectrum.
In the same work~\cite{takeuchi11}, see also Chap.~11 of~\cite{Pikovsky}, 
a partially analytic treatment concluded that the largest Lyapunov exponent (LE) $\lambda$ converged
to its asymptotic value $\lambda_\infty$ logarithmically with the 
system size $N$: 
\begin{equation}
\lambda_\infty -\lambda(N)\simeq\frac{c}{\ln N}.
\label{ln}
\end{equation}
Here $c$ is a positive constant, and the symbol $\simeq$ denotes equality 
after neglecting marginal contributions in $N$. 

In this paper, we study turbulent GCMs with  positive multipliers,
finding that the LE converges to its infinite-size limit 
in a strongly nonuniversal manner. We show that,
depending on the coupling strength and multipliers statistics, 
the LE can follow either a power law
\begin{equation}
    \lambda_{\infty}-\lambda (N) \simeq \frac{c}{N^{\gamma}} \qquad \mbox{with $0<\gamma\le1$} ,
    \label{powerlaw}
\end{equation}
or still a different, arguably slower, convergence law with $N$ in certain situations.

Our results have important implications for turbulent
GCMs with multipliers adopting both signs.
The theoretical approach developed in \cite{takeuchi11}, and claimed to support the scaling law \eqref{ln}, 
did not require any condition on the sign
of the multipliers. Still, for positive multipliers, 
we find a different scaling law, given by Eq.~\eqref{powerlaw}.
We solve this conflict by re-thinking the theoretical 
analysis done by Takeuchi et al.~\cite{takeuchi11} and pointing out a loophole 
in their argumentation. In consequence, the actual asymptotic scaling law of the LE
for general GCMs (i.e.~with positive and negative multipliers)
remains to be rigorously determined.
At the light of our results, even the mere existence of a 
unique scaling law for $\lambda(N)$ turns out to be uncertain.

\section{Globally coupled maps}

The dynamics of a population of $N$
globally coupled one-dimensional maps is iteratively governed by
\begin{eqnarray}
    y_{i}^{t+1} = f\left( (1-\epsilon) \, y_{i}^t
    + \epsilon \bar y^t
    \right), 
    \label{eq:gcm_fx}
\end{eqnarray}
where the index $i \in \{1, \dots, N\}$ labels the $i$-th map and $\bar y^t\equiv N^{-1} \sum_{j=1}^{N} y_{j}^t$ yields the all-to-all coupling. Here $t$ is
a discrete index for time.
For each map, $y_i$ is a scalar variable and 
the nonlinear function $f$ defines the map. $f$ is chosen such that 
yields chaotic dynamics for the uncoupled maps ($\epsilon=0$). 
For small values of the all-to-all coupling $\epsilon$
the GCM \eqref{eq:gcm_fx} displays a fully turbulent phase \cite{kaneko_phd90} 
with $N$ positive Lyapunov exponents.

To calculate the LE  
Eq.~\eqref{eq:gcm_fx} is linearized, 
thereby obtaining the mapping rule for infinitesimal perturbations, 
$v_i^t \equiv \delta y_{i}^t$:
\begin{equation}
    v_{i}^{t+1} = f'\left( (1-\epsilon) y_{i}^t
    + \epsilon \bar y^t
    \right) \times \left[ (1-\epsilon) \, v_i^t+\epsilon \bar v^t\right]
    \label{eq:tangent}
\end{equation}
where $f'$ stands for the derivative of $f$, and
$\bar v^t\equiv N^{-1} \sum_{j=1}^{N} v_{j}^t$.
The time- and site-dependent factors
$f'$ are hereafter referred to as the multipliers of the tangent dynamics.
As time evolves, an arbitrary initial $N$-vector $\bm{v}^0=(v_1^0,\ldots,v_N^0)$
converges to a statistically stationary
configuration $\bm{v}^t$, called
the Lyapunov vector.

The LE is a scalar quantity that characterizes the average exponential growth rate
of infinitesimal perturbations:
$\lambda= \lim_{t\to\infty} \frac{1}t \ln\|\bm{v}^t\|$. 
We may also obtain $\lambda$ averaging the instantaneous logarithmic growth rate
of the Lyapunov vector:
\begin{equation}
\lambda=\left< \ln \left(\frac{\|\bm{v}^{t+1}\|}{\|\bm{v}^t\|}\right)\right> .
\label{LE}
\end{equation}
The bracket denotes the average over an infinite trajectory.
According to the multiplicative Oseledets theorem~\cite{Pikovsky,oseledec}, the value of $\lambda$ is (with probability one) the same 
for all initial perturbations $\bm{v}^{t=0}$, and all orbits 
starting in the basin of attraction of the chaotic attractor, provided the system is 
ergodic.
Moreover, $\lambda$ is an invariant that does not depend on the coordinate system 
nor on the specific norm type used in~\eqref{LE}.

In general the Lyapunov vector components  may fluctuate between positive and negative signs.
However, if $f'$ takes only positive values, then all the Lyapunov
vector components have the same sign (in other words, this is an absorbing configuration).

\begin{figure}
    \centering
    \includegraphics[width=0.9\linewidth]{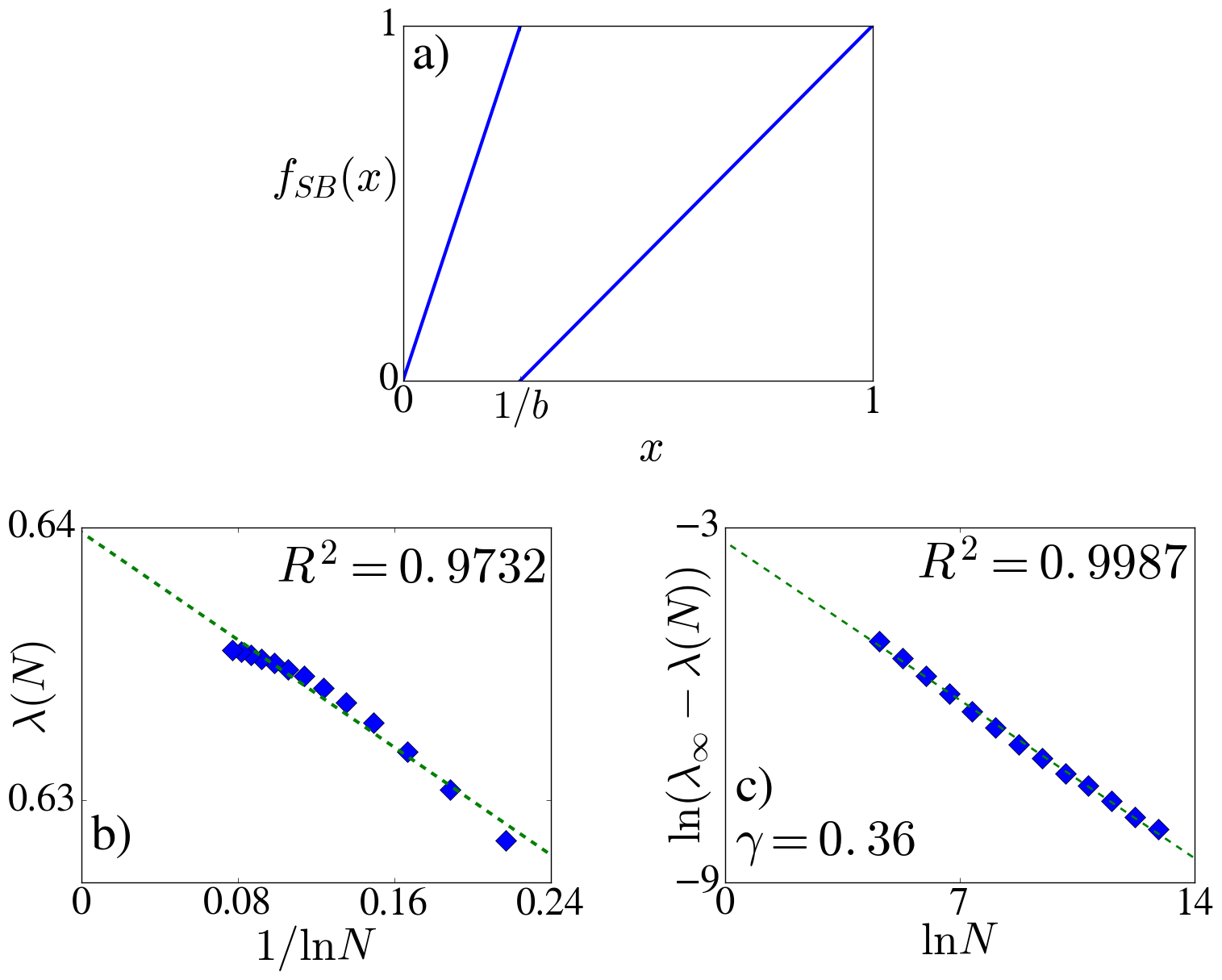}
    \caption{(a) Skewed-Bernoulli map
        \eqref{eq:sb-m} with $b=4$. 
    (b,c) Two alternative forms of representing the LE
    of the SB-GCM  as a function of the system size $N$, see $x$-axis.
Parameters are $\epsilon=0.02$, $b=4$, and the largest size is $N=409600$.
The goodness of the linear fittings is indicated in each panel by the 
regression coefficient $R^2$.}
\label{fig:le-n}
\end{figure}

\section{Preliminary numerical evidence}

As a prototype of map with positive $f'$ (multipliers)
we choose the skewed-Bernoulli (SB) map:
\begin{equation}
    f_{SB}(x) = \left\{ 
        \begin{aligned}
            & b \, x    \hspace{20pt} && \text{if }0\le x\le\frac{1}{b} \\
            & \frac{bx -1}{b-1}&&  \text{if } \frac{1}{b}< x\le1 
        \end{aligned}
    \right.
    \label{eq:sb-m}
\end{equation} 
Parameter $b$ controls the chaoticity of the map. Figure 1(a)
shows $f_{SB}(x)$ for parameter $b=4$.
The LE of a single uncoupled map depends on $b$ as
$\lambda_1=\ln b -(1-1/b)\ln(b-1)$, 
which takes the reference value $\lambda_1=0.5623\ldots$ for $b=4$. 
We adopt $b>2$, since for $b=2$ (skewness-free case) the dynamics is trivially chaotic with no intermittency.

Hereafter, the GCM made up of SB maps is referred to as SB-GCM for abbreviation. The coupling constant $\epsilon$ in Eq.~\eqref{eq:gcm_fx} is chosen small, as this ensures a
fully turbulent dynamics. The reference value $\epsilon=0.02$
was selected in \cite{takeuchi11} and in Chap.~11 of \cite{Pikovsky}. 
The numerical value of the LE for the SB-GCM with $(b,\epsilon)=(4,0.02)$ and different system sizes is represented in Figs.~\ref{fig:le-n}(b) and \ref{fig:le-n}(c). In each plot
a different scaling with $N$ is assumed. In Fig.~\ref{fig:le-n}(b)
we represent $1/\ln N$ in the $x$-axis, and 
a linear fit yields $\lambda_\infty$ and the 
slope $-c$. For comparison in Fig.~\ref{fig:le-n}(c) 
a power-law scaling of the LE, see Eq.~\eqref{powerlaw}, is assumed instead, 
such that the data are fitted to a straight line 
in log-log scale: $\ln[\lambda_\infty-\lambda(N)]= k-\gamma \ln N$.
Our strategy was to determine what value of $\lambda_\infty$ 
yields an optimal linear fit to our data.
For the particular choice of the coupling strength $\epsilon=0.02$ and $b=4$ we
obtain $\gamma=0.36$. 
The fitting is
apparently superior with the power law than with the logarithmic law. 
However, in the former case we have three fitting parameters
instead of two. 

\begin{figure}
    \centering
    \includegraphics[width=\linewidth]{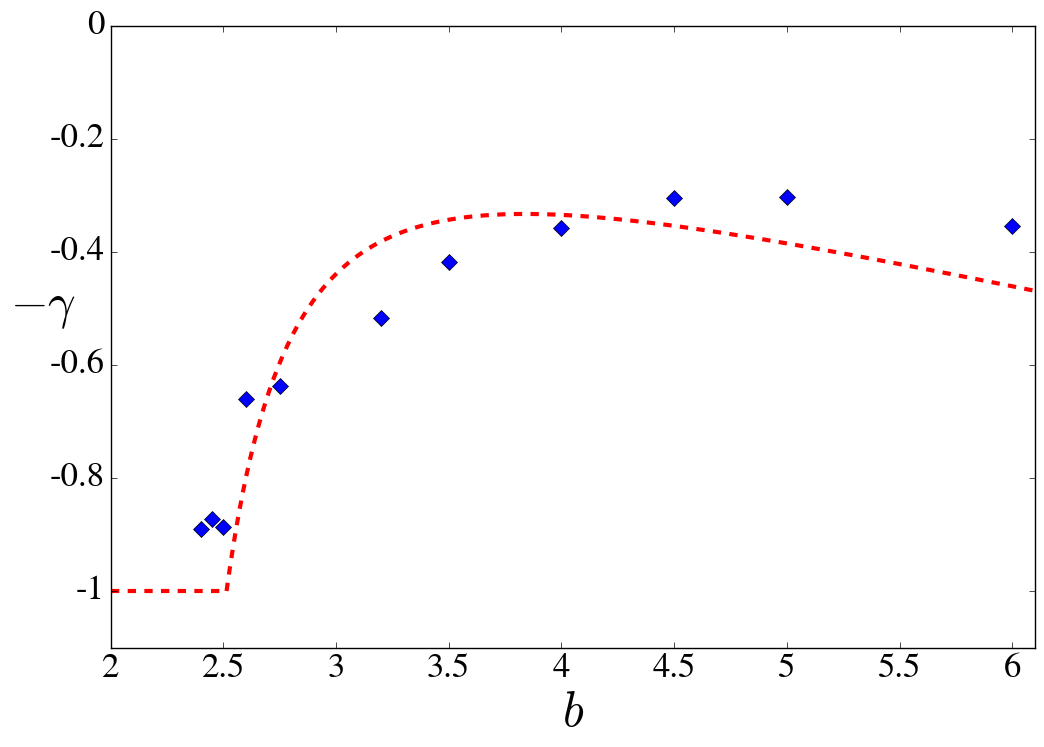}
    \caption{Empirical values of the exponent $-\gamma$ for the SB-GCM
    as a function of the map parameter $b$. The coupling
constant is fixed at $\epsilon=0.02$. The red 
	dashed line corresponds to the theoretical result for the RM model with bi-delta multipliers and the same parameter values.}
    \label{fig:sb-m_b-gamma}
\end{figure}

In order to increase the numerical evidence we measured the LE
for several values of $b$ (fixing $\epsilon=0.02$) and determined 
the exponent $\gamma$ following the procedure outlined above.
Figure \ref{fig:sb-m_b-gamma} shows the 
measured value of $\gamma$ as a function of $b$. A significant
variation in the value of $\gamma$ is apparent.
The main goal of this paper is ascertain the power-law 
convergence of $\lambda(N)$ to $\lambda_\infty$, and explain the dependence
of the exponent $\gamma$ on parameters.

\section{Main Results}

Before presenting our theoretical results, and more
numerical simulations, it is convenient to anticipate which are
the main results of this paper. 
In~\cite{takeuchi11} the logarithmic law~\eqref{ln} was linked
to the power-law tail of the distribution of
Lyapunov vector components. More specifically, in the thermodynamic 
limit the decay was claimed to be an inverse square law: $P(v\gg1)\simeq c/v^2$.
Here, in contrast, we find that the tail can obey a more general expression:
$P(v\gg1)\simeq c/v^{1+\alpha}$, where the tail index $\alpha$ depends
on the model parameters. 
We distinguish three different regimes ---labeled I, II and III---, in which the 
convergence of the LE to $\lambda_\infty$ is different:

\begin{enumerate}
 \item A first regime (I) with $\alpha\ge2$ in which, the LE exhibits a robust
 power law, given by Eq.~\eqref{powerlaw} with exponent $\gamma=1$.
 \item A second regime (II) in which $1<\alpha<2$ and the exponent
 $\gamma$ varies in the range (0,1) with a smooth dependence on the parameters.
 \item A third regime (III), which is only present when the values of the multipliers
are very broadly distributed.
 This regime is much more complicated to analyze in detail, as $\alpha$ 
 takes the value 1 (or smaller). 
 This fact renders the analysis much more cumbersome
 and the main properties of this
 regime remain, in spite of our efforts, largely unknown. 
 Our numerical simulations are consistent with a generalized
 logarithmic scaling $\lambda_\infty-\lambda(N)\simeq(\ln N)^{-\delta}$, but we 
 must be cautions upon drawing general conclusions in this case.
\end{enumerate}

\section{Random multiplier model}

Given that direct numerical results with GCMs will be always inconclusive,
we turn our view to a minimal model that can be 
theoretically analyzed.  This is a
stochastic model of the tangent-space dynamics of GCMs
proposed in Ref.~\cite{takeuchi11}. 
The model simply replaces the local multipliers $f'$ in Eq.~\eqref{eq:tangent}
by independent identically distributed random numbers $\mu_i^t$.
Hence we have
\begin{equation}
    v_{i}^{t+1} = \mu_{i}^t 
    \left[ (1 - \epsilon) v_i^t 
    + \epsilon \bar v^t\right].
    \label{eq:model}
\end{equation}
Note that ignoring correlations between the multipliers is tantamount  to
ignoring the collective dynamics of the mean field $\bar y^t$ present in actual GCMs. 
As in \cite{takeuchi11}, we assume weak correlations  
induced by the collective dynamics
do not alter the final result. Our analytical results are entirely based on the random
multiplier (RM) model~\eqref{eq:model}. 

Before starting the analysis of~\eqref{eq:model}, we briefly
introduce the three multiplier densities used to assess the validity of our results.
Our first case study is the bi-delta density:
\begin{equation}
\rho_{BD}(\mu)=\frac1b\delta(\mu-b)+\frac{b-1}{b}\delta[\mu-b/(b-1)]  .
\label{bidelta}
\end{equation}
This form for $\rho(\mu)$ corresponds to
the binary occurrence of $f'$ for an unperturbed SB map with parameter $b$.

The second example is the log-normal distribution:
\begin{equation}
\rho_{LN}(\mu)=\frac{1}{\sqrt{2\pi} a \mu} e^{-(\ln \mu)^2/(2a^2)} .
\label{log_normal}
\end{equation}
This distribution was originally considered in \cite{takeuchi11}
for the absolute value of the multiplier $|\mu|$.
It is
implemented by taking the exponential of uncorrelated 
zero-mean Gaussian random variables $\xi_{j}^t$:
$\mu_{j}^t = \exp(\xi_j^t)$. The variance of $\xi_j^t$ being $a^2$.

The last case study is the log-uniform distribution, in which
the multipliers are chosen as the exponential 
of a uniform random variable in the interval $[-m,m]$.
The multiplier density in this case has the form
\begin{equation}
    \rho_{LU}(\mu)= \begin{cases}
\frac{1}{2m \mu}  & \mbox{if $e^{-m}<\mu<e^m$}\\
0                & \mbox{otherwise}
               \end{cases}
    \label{log_unif}
\end{equation}

\section{The asymptotic Lyapunov exponent $\lambda_\infty$}
\label{sec:linfty}

Restricting to positive multipliers allows us to 
obtain analytical expressions for the LE.
First, we average both sides of Eq.~\eqref{eq:model}:
\begin{equation}
    \bar{v}^{t+1} =
    (1-\epsilon)\overline{\mu v}^t
    + \epsilon \bar{\mu}^t \bar{v}^t ,
    \label{eq:av}
\end{equation}
where $\overline{\mu v}^t \equiv N^{-1}\sum_{j=1}^N \mu_j^t v_j^t$.
The positiveness of the vector components makes
their average equal to the taxicab norm;
or more formally, $\bar v^t=\|\bm{v}^t\|_1 = (1/N)\sum_{i=1}^{N} v_i^t$. 
From Eq.~\eqref{eq:av} we obtain 
the ratio between consecutive perturbation averages:
\begin{equation}
\frac{\bar{v}^{t+1}}{\bar{v}^t}=\bar\mu^t \left[\epsilon +
(1-\epsilon) \frac{\overline{\mu v}^t}{\bar\mu^t\bar v^t }\right] ,
\end{equation}
and, according to Eq.~\eqref{LE}, 
the LE equals the average of the logarithm of the above formula:
\begin{equation}
 \lambda(N)= \left< \ln  \bar{\mu}^t\right> 
 + \left<\ln
 \left[\epsilon +
(1-\epsilon) \frac{\overline{\mu v}^t}{\bar\mu^t\bar v^t }   \right]\right> .
\label{ln_exact}
\end{equation}

We wish to determine here the value of the LE in the thermodynamic limit
$\lambda_\infty=\lambda(N\to\infty)$. Of the two terms contributing
to $\lambda(N)$ in Eq.~\eqref{ln_exact}, the first one is trivial
since the sample average of the multipliers converges to its mean:
$\lim_{N\to\infty}\bar\mu^t=\langle \mu\rangle$.
To recognize the asymptotic behavior of 
\begin{equation}
s^t\equiv \frac{\overline{\mu v}^t}{\bar\mu^t\bar v^t }
\label{s}
\end{equation}
in Eq.~\eqref{ln_exact} is crucial to complete the result. 

As a preliminary step we prove first that the expected 
value of $s^t$ equals 1,
just assuming the all the Lyapunov vector components are
statistically equivalent. First, we rewrite 
$s^t=\sum_j \tilde \mu_j^t v_j^t/(\sum v_j^t)$, 
where $\tilde \mu_j^t=\mu_j^t/\bar\mu^t$. 
Now the expected value of $s$ is 
\begin{equation}
\langle s^t \rangle= \left< \frac{\sum_j \tilde\mu_j^t v_j^t}{\sum_j v_j^t} \right>
=\left< \sum_{j=1}^N \tilde\mu_j^t v_j^t\right>=
\sum_{j=1}^N \left<  \tilde\mu_j^t\right> \left<v_j^t\right>=1 ,
\end{equation}
where we have chosen the normalization $\sum_j{v_j^t}=1$ in the second equality,
used the independence of $\tilde\mu_j$ and $v_j^t$, and 
substituted $\langle\tilde\mu_j^t\rangle$ and $\langle v_j^t\rangle$ 
by $1$ and $1/N$, respectively.
Now, we consider the thermodynamic limit
of Eq.~\eqref{ln_exact}. Due to the convexity of the logarithm,
the expected value of 
$\ln[\epsilon+(1-\epsilon) s^t]$
is not larger than 
$\ln[\epsilon+(1-\epsilon) \langle s^t \rangle]$,
we obtain that 
\begin{equation}
\lambda_\infty \le \ln\langle\mu\rangle .
\label{cons}
\end{equation}

The previous constraint turns into an equality with a few additional assumptions.
Let us assume that, given a certain norm (e.g. $\bar v=1$), 
each Lyapunov vector component is 
uncorrelated from the rest and it is drawn from a stationary
probability density $P_s(v)$. If such a $P_s(v)$ really exists is discussed later on.
As the multipliers and the vector components are uncorrelated,
we have a quite robust trivial result: $\lim_{N\to\infty}s^t=1$, 
provided the expected value of $v$ exists, see e.g.~\cite{cohn82}. 
We get thus the simple relation:
\begin{equation}
\lambda_\infty=\ln\langle\mu\rangle .
\label{linfty}
\end{equation}
This value of $\lambda_\infty$ is larger than $\lambda_1=\langle\ln\mu\rangle$, 
the LE for a single uncoupled map. This means that an extreme `coupling sensitivity of
chaos' \cite{daido84} shows up in the thermodynamic limit.
The identity in Eq.~\eqref{linfty} is valuable for
the numerical validation of the theory with the RM model since $\lambda_\infty$
is not a fitting parameter anymore (in contradistinction to the general case of GCMs).

\section{The Lyapunov vector and its localization}

Before addressing our main question (i.e.,~the size dependence 
of the LE), it is necessary to suitably describe the Lyapunov vector in the thermodynamic limit ($N\to\infty$). 
Indeed, as shown later, the localization strength of the Lyapunov vector is 
intimately related with the convergence of the LE with $N$.

First of all, we note that in the thermodynamic limit 
Lyapunov vector components
are expected to 
be distributed as a stationary density 
if the exponential amplification is removed~\cite{Pikovsky}: 
$P(v,t)=P(v \, e^{-\lambda_\infty t},0)$
\footnote{This is tantamount assuming that the diffusion coefficient
accompanying chaotic amplification vanishes in the large size limit. 
This is a plausible assumption, since so far this has been found to be violated 
only for some Hamiltonian lattices \cite{plp16}.}.
The norm of the Lyapunov vector is irrelevant as it can always be scaled out.
Hence we only need $P(v,0)=P_s(v)$.
Takeuchi et al.~\cite{takeuchi11} addressed the problem
resorting to a Hopf-Cole transformation and then solving the stationary solution
of Fokker-Planck equation.
We replicate part of their mathematical treatment in the following lines.
We note, however, that the conclusions of our analysis are radically different.

First of all, we make the Hopf-Cole transformation of the vector components:
\begin{equation}
u_i^t=\ln v_i^t.
\label{HopfCole}
\end{equation}
As we are assuming positive multipliers $\mu_j^t>0$, 
the $v_i^t$ remain above zero at all times, hence
no absolute value is required to take the logarithm
\footnote{In~\cite{takeuchi11}, the transformation $u_i^t=\ln|v_i^t|$
is taken without paying much attention to the absolute value. This is not completely unreasonable if one considers that in spatio-temporal chaos 
the absolute value causes no effect in the universality class and 
thereupon the associated critical exponents \cite{pik98}.}. 

In terms of the $u$ variables the evolution equation of the RM model \eqref{eq:model} becomes:
\begin{equation}
u_j^{t+1}=u_j^t+\ln\mu_j^t+ \ln(1-\epsilon) 
+  \ln\left(1+\frac{\epsilon \bar v^t e^{-u_j^t}}{1-\epsilon}  \right).
\end{equation}
For simplicity of notation, we keep $\bar v^t$ instead of writing $\overline{e^{u}}^t$.

Replacing a discrete difference in time by a time derivative,
the corresponding Fokker-Planck equation
for the density $\tilde P(u,t)$ 
in the co-moving reference frame 
at velocity $\lambda_\infty$ is
\begin{equation}
\partial_t \tilde P(u,t)=-\partial_u[(\lambda_0(u)-\lambda_\infty)\tilde P(u,t)]+\frac{D}{2}\partial_{uu} \tilde P(u,t),
\label{eq:fp}
\end{equation}
where
\begin{equation}
\lambda_0(u)= \langle\ln \mu\rangle +\ln(1-\epsilon)+ \ln\left(1+\frac{\epsilon \bar v e^{-u}}{1-\epsilon} \right) ,
\label{l0}
\end{equation}
and the constant $D$ is the variance of the noise, which is given by
\begin{equation}
D=\mathrm{var}(\ln\mu)
\label{d}
\end{equation}
If the constant $\lambda_\infty$ is the LE in the limit $N\to\infty$, a stationary solution, $\tilde P_s(u)$, of~\eqref{eq:fp}
exists and is given by the solution of 
\begin{equation}
\frac{d}{du}[(\lambda_\infty-\lambda_0(u))\tilde P_s(u)]+\frac{D}{2}\frac{d^2}{du^2} \tilde P_s(u)=0.
\label{stationary}
\end{equation}
We do not need the exact solution of this equation, 
only the asymptotic (large $u$) decay of $\tilde P_s(u)$ will be of our interest.
The solution of~\eqref{stationary} exhibits an exponential decay:
\begin{equation}
\tilde P_s(u\to\infty)\simeq k \, e^{-\alpha u},
\label{psu}
\end{equation}
where $\alpha=2 [\lambda_\infty-\lambda_0(u\to\infty)]/D$ measures 
the Lyapunov vector localization strength.
Recalling Eqs.~\eqref{l0} and~\eqref{d} we can express $\alpha$ in terms
of $\epsilon$ and the statistical properties of $\mu$:
\begin{equation}
\alpha=2\,\frac{\lambda_\infty-\langle\ln \mu\rangle-\ln(1-\epsilon)}{\mathrm{var}(\ln\mu)} . 
\label{alpha}
\end{equation}
This formula relates the Lyapunov vector localization index $\alpha$ 
with $\lambda_\infty$, 
the multiplier density, and the coupling strength $\epsilon$. As intuitively expected $\alpha$ grows with $\epsilon$, i.e., the vector
becomes less localized as the coupling is increased.

\begin{figure}
    \centering
    \includegraphics[width=0.9\linewidth]{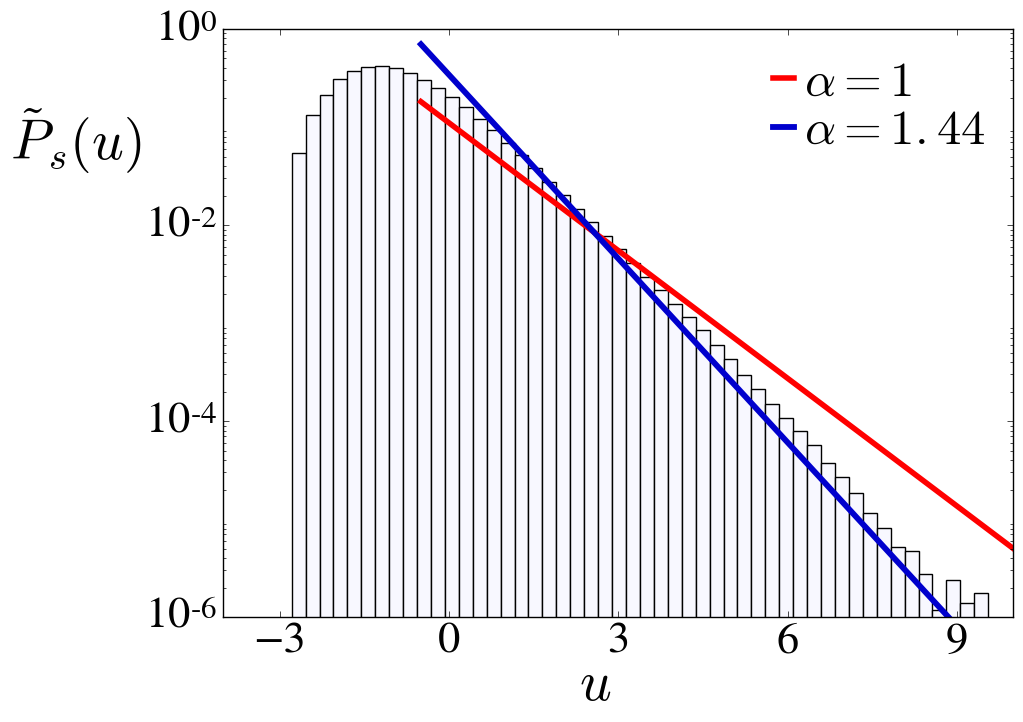}
    \caption{Probability density 
    $\tilde P_s(u)$ obtained form numerical simulations of the RM model with bi-delta density~\eqref{bidelta}
and parameters $\epsilon=0.02$, $b=3$.
The system size is $N=409600$, and the
components of the Lyapunov vector were 
retrieved at 100 different times, fixing $\bar v=1$, to estimate $\tilde P_s(u)$. 
The straight lines correspond to the exponential $\propto e^{-\alpha u}$, with  $\alpha=1$ (red), as predicted in~\cite{takeuchi11},
and $\alpha=1.44$ (blue), as obtained from 
our theory in Eq.~\eqref{alpha_c}, see also Table I.}
\label{fig:Density_Pu}
\end{figure}

Moreover, as we already know $\lambda_\infty$, via Eq.~\eqref{linfty},
the value of exponent $\alpha$ in Eq.~\eqref{alpha}
becomes completely determined:
\begin{equation}
\alpha=2\,\frac{\ln\langle\mu\rangle-\langle\ln \mu\rangle-\ln(1-\epsilon)}{\mathrm{var}(\ln\mu)} .
\label{alpha_c}
\end{equation}
This formula relates the localization strength of the Lyapunov vector,
i.e.~its tail index $\alpha$, with known quantities.
As a numerical check, the empirical $\tilde P_s(u)$ is represented in Fig.~\ref{fig:Density_Pu} for specific parameters of
the RM model with bi-delta multiplier density.
The observed decay rate at large $u$ is in good agreement with the result of Eq.~\eqref{alpha_c}: $\alpha=1.44$.
The asymptotic slope predicted by Takeuchi et al.~is $-2$ (i.e.,~$\alpha=1$),
which is in clear
disagreement with the data.

\begin{figure*}
\includegraphics[width=0.9\textwidth]{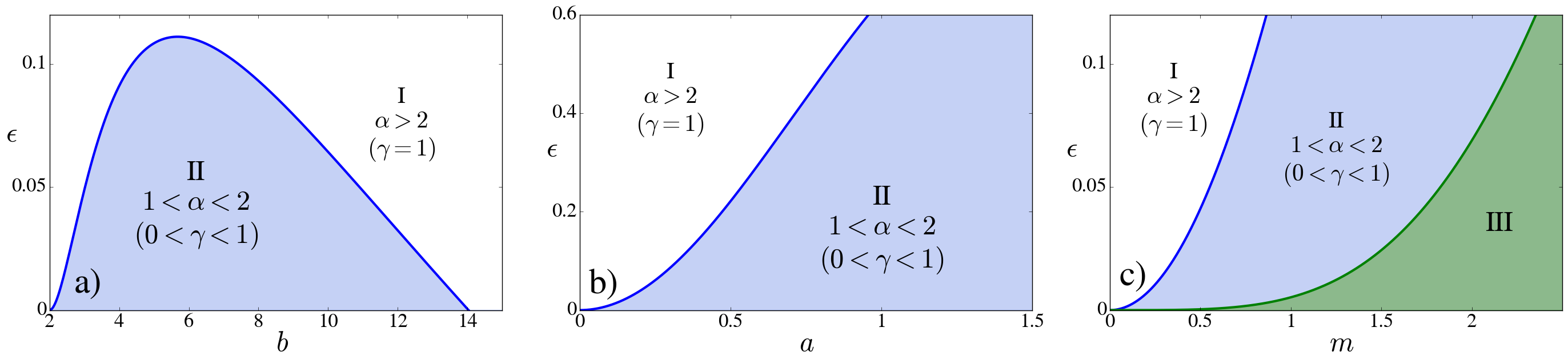}
\caption{Phase diagrams
of the RM model for three multiplier densities: (a) Bi-delta, Eq.~\eqref{bidelta}; (b) Log-normal, Eq.~\eqref{log_normal}; (c) Log-uniform, Eq.~\eqref{log_unif}.
Level lines of $\alpha=1,2$,
obtained from Eq.~\eqref{alpha_c},
enclose regimes I, II and III.}
\label{phase_diag}
\end{figure*}

Reverting the Hopf-Cole transformation in~\eqref{HopfCole}, Eq.~\eqref{psu}
translates into a power-law tail of the stationary density of the vector components:
\begin{equation}
P_s(v\to\infty)\simeq k'\,  v^{-1-\alpha} .
\label{ps}
\end{equation}
We can go one step forward and use Eq.~\eqref{alpha_c} to generate
phase diagrams for the three multiplier probability distribution types
introduced in Sec.~V. 
In Fig.~\ref{phase_diag} we show the phase diagrams for the bi-delta, log-normal,
and log-uniform multiplier densities in panels (a), (b) and (c), respectively.
The ranges of $\epsilon$ depicted in Fig.~\ref{phase_diag} can be particularly 
large since the RM model is always representing the turbulent regime.
In actual GCMs, turbulence typically ceases to exist already for $\epsilon\sim 0.1$.

The level lines $\alpha=2$ and $\alpha=1$ are specially interesting,
as they mark the boundaries between different regimes, 
depending on the statistical properties of the random multipliers and coupling strength. In particular,
the line $\alpha=2$, Fig.~\ref{phase_diag}(c), 
is the boundary that separates models for which the 
probability density of vector components, $P_s(v)$, exhibits a finite variance from those where it does not.
The effect of model parameters on the asymptotic scaling properties of the LE with the system size will be 
analyzed in detail in Sec.~\ref{sec:regimes}.

The existence of these boundary lines depends on the random multipliers specific 
statistics. Note, for instance, that the level line $\alpha=1$, does 
not exist for the bi-delta
multiplier distribution, and coincides with $\epsilon=0$ for the 
log-normal distribution (see Figs.~\ref{phase_diag}(a) and \ref{phase_diag}(b)). 
However, for the log-uniform multiplier distribution
this line is indeed present at finite $\epsilon$ values, see Fig.~\ref{phase_diag}(c).
We stress here that, in the green shaded region, Fig.~\ref{phase_diag}(c), 
our previous theory breaks down, since
it predicts $\alpha<1$. This possibility is forbidden because
for any finite population, such a density leads to a paradoxical
result, as we explain in the following. 
In a finite population we can fix $\bar v=1$, and
the largest component $v_{max}$ cannot be larger than $N$. However,
if one draws the vector components from a density with a tail
decaying as $v^{-1-\alpha}$ inconsistencies arise.
The probability for one vector component being larger than $N$ is
$\Pr(v>v_{max}=N)=\int_N^{\infty} P_s(v) dv\sim N^{-\alpha}$;
and the probability all components are below $N$ is roughly
$(1-N^{-\alpha})^{N}\simeq1-N^{1-\alpha}$, which approaches 1 if $\alpha>1$.
If $\alpha<1$,
some components will exceed the largest allowed value $v_{max}=N$ almost
surely as $N$ grows. 
In the marginal case $\alpha=1$, the situation is exactly at the edge.

In the next section we analyze the regular case, $\alpha>1$,
and derive a power-law convergence for the LE. 
The study of the anomalous green region in Fig.~\ref{phase_diag}(c), 
is postponed to Sec.~\ref{r3}.

\begin{table*}
\caption{Statistical properties of the three multiplier 
probability density types we study in this paper. 
The last column is the theoretical prediction for the tail index
$\alpha$ describing the asymptotic decay of the Lyapunov vector 
components distribution for each RM model, according to Eq.~\eqref{alpha_c}. (*) 
In the case of the log-uniform density the result is valid only for $\alpha>1$.}
\centering 
\begin{tabular}{c c  c  c  c } 
\hline\hline 
Density $\rho(\mu)$ & $\lambda_\infty=\ln\langle\mu\rangle$ &  $\langle \ln\mu \rangle$ & $\mathrm{var}(\ln\mu)$ & Tail index$^*$ $\alpha$ \\ 
\hline \\ [-1ex]
Bi-delta, Eq.~\eqref{bidelta} &  $\ln 2$ &  $\frac1b \ln b + \frac{b-1}{b} \ln\left(\frac{b}{b-1}\right)$ &  
$\frac{(b-1) \ln^2(b-1)}{b^2}$ &  
$\frac{2 b \left[b \ln \left(\frac2{1-\epsilon }\right)-(b-1) \ln \left(\frac{b}{b-1}\right)-\ln b\right]}{(b-1) \ln^2(b-1)}$\\  [1ex]  
Log-normal, Eq.~\eqref{log_normal}  & $\frac{a^2}{2}$ & $0$   & $a^2$ & $1-\frac{2\ln(1-\epsilon)}{a^2}$ \\[1ex] 
Log-uniform, Eq.~\eqref{log_unif}  & 
$
\ln\left(\frac{\sinh m}m\right) 
$
& $0$   & $\frac{m^2}3$ 
& 
$
\frac{6 \left[\ln \left(\frac{\sinh m }{m }\right)-\ln (1-\epsilon )\right]}{m^2} 
$
\\[1ex] 
\hline\hline 
\end{tabular}
\label{table} 
\end{table*}

\section{Regimes I  and II: Power-law convergence of the Lyapunov exponent}
\label{sec:regimes}

The convergence of the LE to $\lambda_\infty$ whenever $\alpha>1$ is analyzed next,
performing a perturbation expansion of Eq.~\eqref{ln_exact}.
We start replacing all $\bar\mu^t$ by $\langle\mu\rangle$ in Eq.~\eqref{ln_exact}. 
This approximation is sensible provided that the multipliers are not
fat-tailed distributed, what we forbid. 
Time fluctuations of $\bar\mu^t$ yield deviations of order $N^{-1}$, 
e.g.~$\ln\bar{\mu}^t\simeq\ln \langle\mu\rangle + O(N^{-1})$. 
These $O(N^{-1})$ terms can be safely
neglected, as the convergence of $\lambda(N)$ to $\lambda_\infty$
is dominated by the statistics of the vector components as verified {\sl a posteriori}.
Keeping in mind that terms of order $O(N^{-1})$ are neglected, Eq.~\eqref{ln_exact} yields
the approximation
\begin{equation}
 \lambda(N)\simeq \lambda_\infty+ \left<\ln\left[\epsilon + (1-\epsilon) s^t \right]\right> .
 \label{init}
 \end{equation}
Now, given that $\lim_{N\to\infty}s^t=1$ and $\langle s^t\rangle=1$, we Taylor expand the logarithm up to second order:
 \begin{equation}
 \lambda_\infty-\lambda(N)\simeq 
 \frac{(1-\epsilon)^2}2 \mathrm{var}( s^t ) .
 \label{logexpand}
 \end{equation}
To proceed further with the  
calculation we can resort to Eq.~(3) 
in Ref.~\cite{cohn82}. Nevertheless, the interested reader can
find the detailed calculation in the
Appendix. The final result for 
the leading order correction to $\lambda_\infty$ is:
\begin{equation}
\lambda_\infty- \lambda(N) \simeq \frac{(1-\epsilon)^2 \mathrm{var}(\mu)}{2 \langle\mu\rangle^2}  
\left<  \frac{\sum_{j=1}^{N} (v_{j}^t)^2}{\left(\sum_{j=1}^{N} v_{j}^t   \right)^2}\right> .
\label{final}
\end{equation}
In the Appendix the goodness of the approximations are tested for
the bi-modal multiplier density. 
Unfortunately, even
for moderate values of $b$, say above 4,
to achieve the asymptotic regime is computationally too demanding
for our current numerical capabilities. 

With Eq.~\eqref{final} the problem reduces to properly estimate the average
in the right hand side. Writing that equation in this form:
\begin{equation}
\lambda_\infty- \lambda(N) \simeq \frac{(1-\epsilon)^2 \mathrm{var}(\mu)}{2 \langle\mu\rangle^2 N}  
\left<  \frac{ {\overline{v^2}}^t  }{ \left(\bar v^t\right)^2}\right> ,
\label{finalbf}
\end{equation}
it becomes apparent that the expected convergence rate 
of the LE would be $N^{-1}$ 
if the Lyapunov vector was completely delocalized, 
i.e.,~all components taking comparable values on average. 
However this is not the case because, as seen above, the components of the Lyapunov vector are distributed with a power-law tail (in the thermodynamic limit). Therefore, we must examine the average in Eq.~\eqref{final} more carefully.

\subsection{Exponent $\gamma$}

Actually, calculating the average 
\begin{equation}
T_N\equiv\left< \frac{\sum_{j=1}^{N} (v_{j}^t)^2}{\left(\sum_{j=1}^{N} v_{j}^t   \right)^2}\right>
\label{tn_av}
\end{equation}
that appears in Eq.~\eqref{final} turns out to be a formidable task. 
We have $N$ non-independent Lyapunov vector components evolving in time.
To proceed further we assume
that the distribution of $v$ in the thermodynamic limit
is all we need to estimate Eq.~\eqref{tn_av} at leading order. Correlations
originating from
the finiteness of the population are regarded as higher-order
effects, which we shall ignore within our approximation.
Thus, we assume vector components $v_j$ are
independent (identically distributed) random variables.
Under this natural assumption analytical results are available 
in the mathematical literature. 
For distributions with a Pareto-type decay and tail index $\alpha$, 
i.e.~Eq.~\eqref{ps}, the asymptotic dependence of $T_N$ on $N$ is analytically known to 
scale as~\cite{mcleish82,albrecher07}:
\begin{equation}
T_N\sim
\begin{cases}
\Gamma(2-\alpha) \ell(N) N^{1-\alpha}  &\qquad \mbox{for $1<\alpha<2$} \\
\langle v^2\rangle N^{-1} & \qquad  \mbox{for $\alpha>2$}
\end{cases}
\label{tn}
\end{equation}
Here $\ell(N)$ is a ``slowly varying function'' satisfying $\lim_{N\to\infty} \ell(N)/\ln(N)=0$, and we have fixed $\langle v\rangle=1$ to make the expressions less convoluted. According to Eq.~\eqref{tn}, the statistic $T_N$ decays as
a power of $N$ for all $\alpha$ values, and so does $\lambda_\infty-\lambda(N)$ by virtue of Eq.~\eqref{final}.
Specifically, if  $\alpha>2$, the probability $P_s(v)$ in~Eq.~\eqref{ps}
has finite variance and the trivial exponent, 
corresponding to a delocalized vector, is immediately recovered.
In contrast, if $1<\alpha<2$ the exponent adopts a nontrivial value: $\gamma = \alpha -1$.
For the sake of clarity, we find it convenient to cast these results 
into a single expression:   
\begin{equation}
\gamma_{i.i.d.}=\min(\alpha-1,1) ,
\label{gamma}
\end{equation}
where $\alpha>1$ has a known dependence on the 
distribution of the multipliers given by our theory through Eq.~\eqref{alpha_c}.
The subscript $i.i.d.$ indicates that the hypothesis of independent vector components
is assumed. 
Remarkably,
the stronger the localization of the Lyapunov vector the slower the
convergence of the Lyapunov exponent, i.e.~$\gamma\to0^+$ as $\alpha\to1^+$. 

\begin{figure*}
\includegraphics[width=0.9\textwidth]{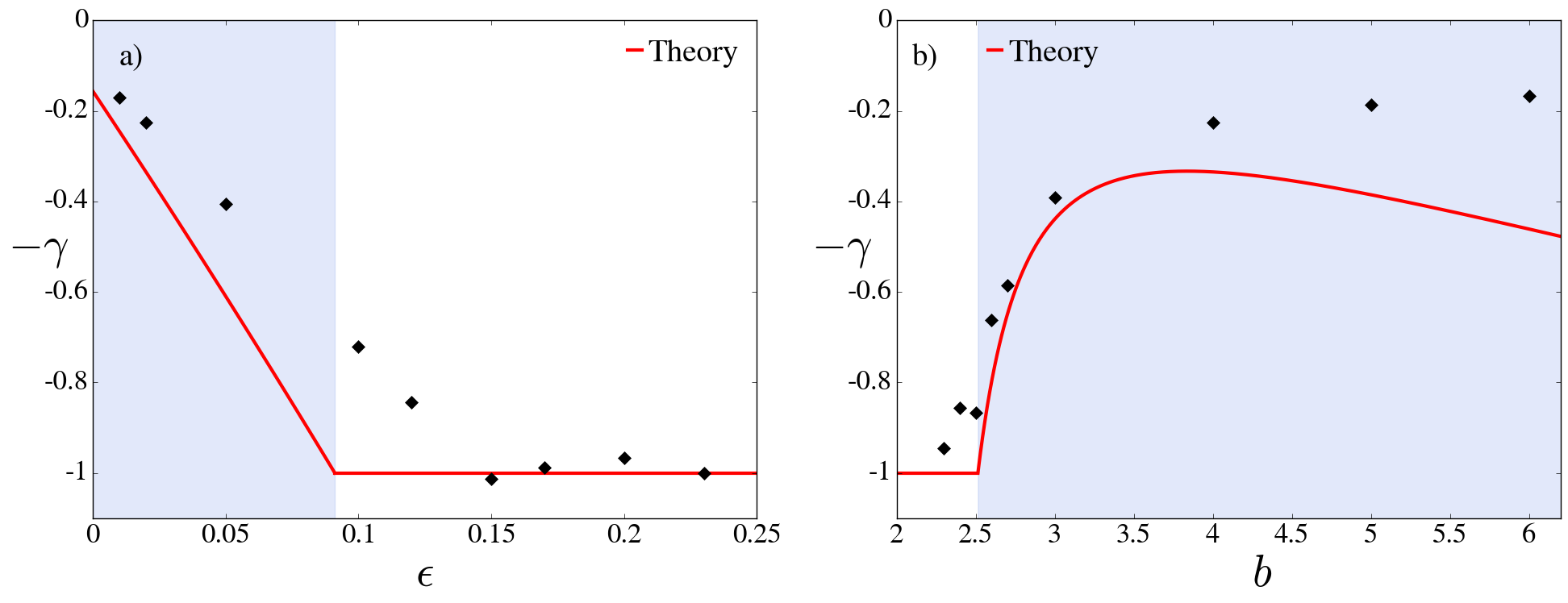}
\caption{ RM with bi-delta density. Numerical estimate of the power-law exponent 
$-\gamma$ and comparison with the theory, Eqs.~\eqref{alpha_c}, and \eqref{gamma}, 
see also Table \ref{table} for $b=4$ in (a), and $\epsilon=0.02$ in (b).
The background shading in both panels indicates parameter values
inside region II of Fig.~\ref{phase_diag}(a).}
\label{num_rmbd}
\end{figure*}

Here concludes the proof of our main result in Eq.~\eqref{powerlaw}, supplemented
by Eqs.~\eqref{gamma} and~\eqref{alpha_c}.
The correctness of our prediction for 
the exponent $\gamma$ relies on the validity of the assumption of
complete independence of the vector components. This is a reasonable
approximation, albeit not fully justified.
Nonetheless, the main result summarized in the power law in Eq.~\eqref{powerlaw} 
is probably quite robust.
For comparison, let us mention that the behavior of $T_N$ when the $v_j$'s 
are drawn in a deterministic way ---selecting values at which 
the cumulative distribution function equals $(j-1)/N$--- is also a power law
with a slightly different exponent~\cite{mcleish82}: $\gamma_{\mathrm{det}}=\min(2-2/\alpha,1)$.
As $\gamma_{i.i.d.}$, also $\gamma_{\mathrm{det}}$ equals 1 for 
$\alpha\ge2$ and vanishes as $\alpha\to1$.
The difference between $\gamma_{\mathrm{det}}$ and $\gamma_{i.i.d.}$ is no more than $0.17$ (the maximum difference is achieved at $\alpha=\sqrt{2}$). 
Expressing some caution, we believe 
this may give an idea of the degree of accuracy of the results based 
upon the $i.i.d.$ hypothesis above.

\subsection{Numerical results}

In this section we test the validity of our results for the RM model
with the three multipliers density types summarized in Table~\ref{table}.
Actually, a thorough numerical verification of the predicted phase diagrams 
in Fig.~\ref{phase_diag} is far too demanding. Alternatively, 
we can determine numerically the exponent $\gamma$ 
along selected sections of the phase diagrams. For specific parameter values, 
$\lambda(N)$ is measured for several system sizes (up to $N=409600$),
and the value of $-\gamma$ is the slope obtained from the linear fit
$\ln[\lambda_\infty-\lambda(N)]= k-\gamma \ln N$, where
$\lambda_\infty$ is known to be $\ln\langle\mu\rangle$.

\subsubsection{Bi-delta multiplier density} 

Irrespective of the particular values of $b$ and $\epsilon$, the Lyapunov
exponent converges to $\lambda_\infty=\ln\langle \mu \rangle=\ln 2$. 
In Figs.~\ref{num_rmbd}(a) and~\ref{num_rmbd}(b), the numerical
estimations of $-\gamma$ are represented at fixed $b$ and fixed $\epsilon$,
respectively. For comparison, the theoretical prediction of $\gamma$, 
via Eq.~\eqref{gamma}, and the $\alpha$ value in Table~\ref{table}
are  plotted as solid lines.
As can be seen in both panels of Fig.~\ref{num_rmbd}, the exponent $\gamma$ 
strongly depends on parameters. In panel (a) the elbow at $\epsilon\approx0.09$
is not accurately captured by the data, but the general behavior of $\gamma$ is
successfully reproduced. In Fig.~\ref{num_rmbd}(b), $\gamma$ 
significantly  departs from 
the theory as $b$ grows above 4. This is not surprising, as our
numerical tests--- see Fig.~\ref{fig:tests} in the Appendix---
already revealed the slow convergence to the asymptotic 
regime for moderate $b$ values.
In any case Fig.~\ref{num_rmbd}(b)  exhibits a trend of $\gamma$
similar to the SB-GCM in Fig.~\ref{fig:sb-m_b-gamma} with the
same $\epsilon$ value.
In our view, this confirms the validity our analysis.

\subsubsection{Log-normal multiplier density}

\begin{figure}
    \centering
    \includegraphics[width=0.9\linewidth]{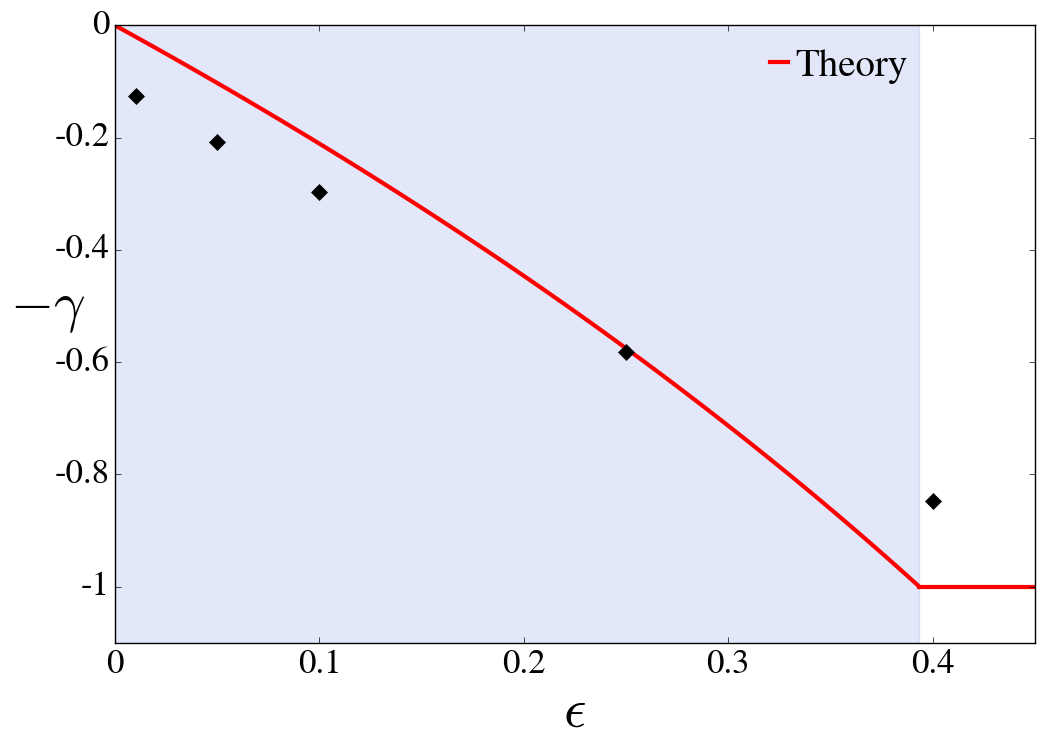}
    \caption{{ RM model with log-normal density with $a=1$. Numerical determination of the exponent $\gamma$ as the coupling $\epsilon$ is varied and comparison with our theoretical prediction.} }
    \label{fig:expgauss-rm_eps-gamma}
\end{figure}

In Fig.~\ref{fig:expgauss-rm_eps-gamma}, we monitor $\gamma$
as a function of the coupling parameter for $a=1$. The numerical
results and theory are in reasonable agreement, in our opinion,
and the general trend of $\gamma$ is fairly reproduced.
As a side note, we point out that including the  exact value of
$\lambda_\infty=\ln\langle\mu\rangle$ in the fittings is crucial. 
The actual values of the multipliers are affected
by the accuracy of the Gaussian random number generator, 
such that the difference between the numerical value of $\ln\langle\mu\rangle$
and the expected value $a^2/2=1/2$ is of the order of $10^{-3}$. 

\subsubsection{Log-uniform multiplier density}

\begin{figure}
    \centering
    \includegraphics[width=0.9\linewidth]{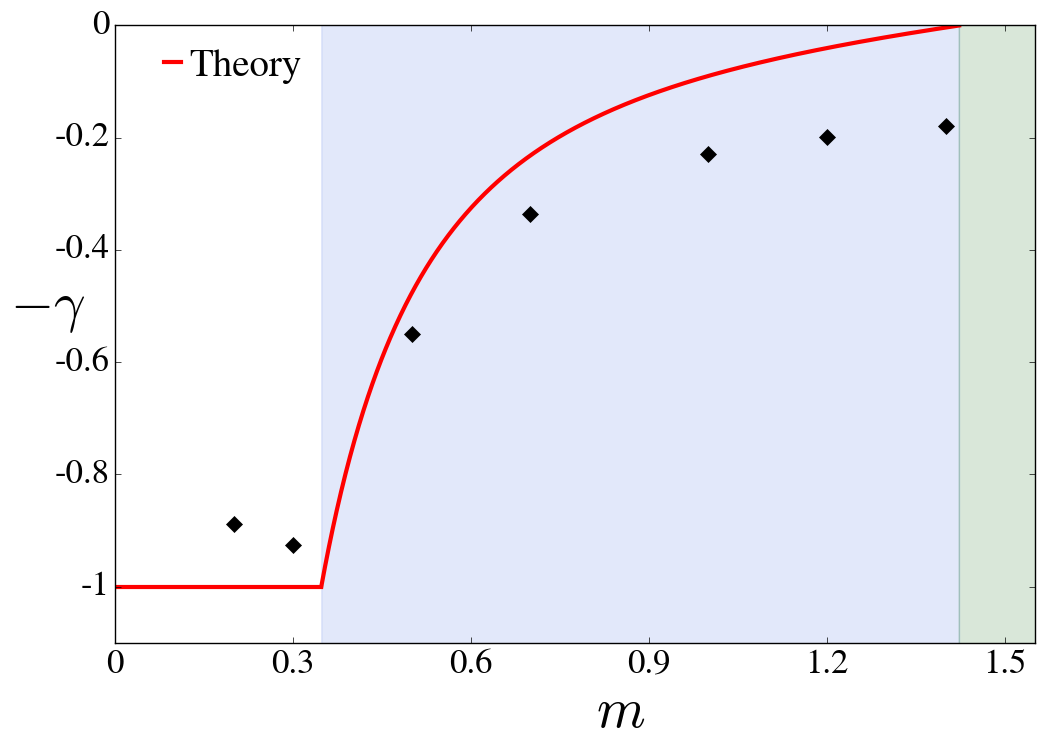}
    \caption{RM model with log-uniform density. Numerical results for
    the exponent $\gamma$ as a function of parameter $m$ for a coupling
    strength $\epsilon=0.02$.
    The background shading indicates parameter values
inside regions II (blue) and III (green) of Fig.~\ref{phase_diag}(c).}
    \label{fig:log_unif_gamma}
\end{figure}

Figure~\ref{fig:log_unif_gamma}  shows the empirical 
values of $\gamma$ as a function of $m$ for $\epsilon=0.02$. As $m$ grows, 
the values of the { random multipliers become increasingly scattered}
and, as already discussed, the boundary of the power-law behavior 
(corresponding to $\alpha=1$) is located 
at the critical value $m_c=1.422\ldots$. 
As { anticipated}, strong finite-size effects in the simulations hinder
the convergence of $\gamma$ to zero, as $m$ approaches $m_c$.
Said that, we estimate the theory { works reasonably well, given the complexity
of the problem}. 

\section{Regime III}
\label{r3}

The goal of describing the Lyapunov dynamics in regime III
requires a non trivial refinement of our theory. 
We advance that theoretical and computational
obstacles have not allowed us to accomplish that goal so far.
Nonetheless, it may be instructive
to devote this section to enumerate the main difficulties
we have encountered, and to discuss some partial results.

In contrast to regions I and II, in region III the components of the Lyapunov vector 
are strongly scattered, such that
the limit $v_{max}$ proportional to $N$ cannot be ignored. 
Roughly speaking, the finiteness of the system is always relevant, and 
it is not even  obvious if a density $P_s(v)$ is meaningful
in the thermodynamic limit. 

\subsection{Vanishing diffusion coefficient}

\begin{figure}
    \centering
    \includegraphics[width=0.9\linewidth]{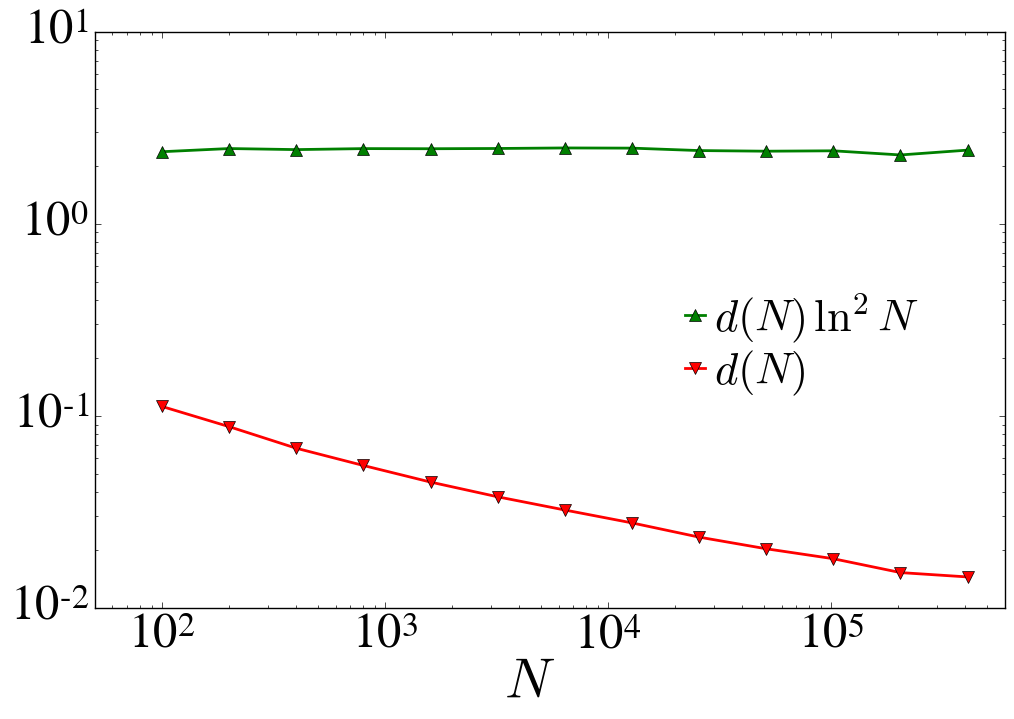}
    \caption{{ Numerical results for the} diffusion coefficient defined 
    by { Eq.~\eqref{diff}} as a function of $N$
    for the RM model with log-uniform multiplier density
    and parameters $\epsilon=0.02$ and $m=2$.}
    \label{fig:dif}
\end{figure}

In order to confirm that the chaotic dynamics is self-averaging in the
thermodynamic limit we computed the diffusion coefficient, characterizing the
intermittency of chaos~\cite{fujisaka84,Pikovsky}. 
Before introducing our numerical results we briefly
summarize a few basic notions. Let $\lambda(\tau,t_0;N)$ be the finite-time Lyapunov exponent of a system of size $N$. 
\begin{equation}
\lambda(\tau,t_0;N)=\frac{1}{\tau}\ln\frac{\bar v^{t_0+\tau}}{\bar v^{t_0}} .
\end{equation}
This quantity depends on the time interval $\tau$, and
on the state of the system through $t_0$. 
The LE is recovered in the limit $\lambda(N)=\lim_{\tau\to\infty}\lambda(\tau,t_0;N)$.
The diffusion coefficient $d$ is an invariant 
that quantifies mean quadratic deviations
from the average exponential growth of infinitesimal perturbations
\cite{Pikovsky}:
\begin{equation}
d(N)=\lim_{\tau\to\infty}
\frac{\left<(\lambda(\tau,t_0;N) \tau-\lambda(N)\tau)^2\right>}{\tau}
\label{diff}
\end{equation}
In a generic chaotic system $d$ is nonzero. In our RM model,
$d$ departs from zero, due to the fluctuations of the FTLE 
caused by $\bar\mu^t$ and $s^t$, see Eq.~\eqref{ln_exact}.
To ascertain { whether} this fluctuation persists in the thermodynamic
limit, we computed $d(N)$
for several system sizes and fixed parameter values
well inside region III ($m=2$ and $\epsilon=0.02$).
The numerical result in Fig.~\ref{fig:dif} shows that the decay of $d(N)$ 
to zero is consistent with 
the inverse of the logarithm squared: $d(N)\simeq c/\ln^2N$. 
To our surprise, this decay is robust over several decades.
The fact that $d(N\to\infty)\to0$ implies that
there exist a co-moving reference frame
in which the Lyapunov vector is stationary
in the thermodynamic limit.

\subsection{The Lyapunov vector}

\begin{figure}
    \centering
    \includegraphics[width=0.9\linewidth]{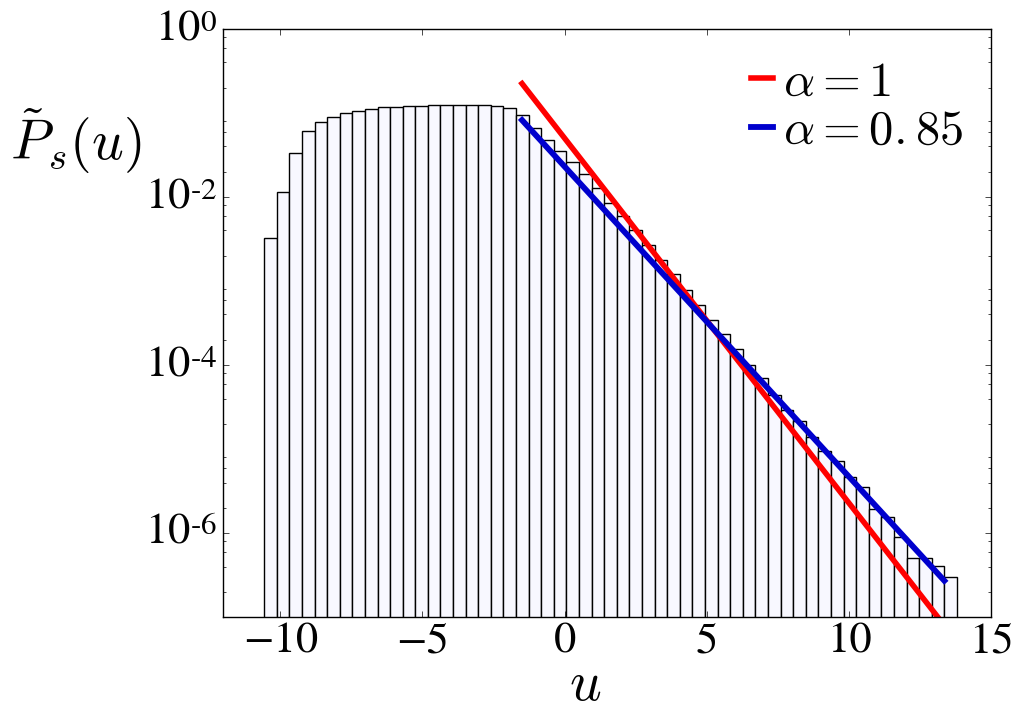}
    \caption{{ Numerical probability distribution of log-transformed vector components $u_j=\ln v_j$ for the log-uniform RM model for} a population of size 
    $N=2^{14}\times100\approx 1.6\times10^6$. Parameters are $m=4$ and $\epsilon=0.02$.
    The histogram is an average over 50 vector configurations with $\bar v=1$.
    The red and blue straight lines are exponentials $\propto e^{-\alpha u}$ 
    with $\alpha=1$ and $0.85$,
    respectively.}
    \label{fig:p}
\end{figure}

In view that a stationary density $\tilde P_s(u)$ exists in the 
thermodynamic limit, we decided to measure it numerically
for a large system size. In Fig.~\ref{fig:p} we can see the distribution
of the (log-transformed) vector components from an average over 50 states.
Notably, the tail decays with a slope that { gives approximately} $\alpha=0.85$, 
which is appreciably steeper than the prediction from Eq.~\eqref{alpha_c}: $\alpha=0.7276\ldots$. As previously discussed, values of $\alpha$ below
1 eventually yield too large vector components, above the maximal allowed
value $u_{max}=\ln N$ (for $\bar v=1$). For the example, in Fig.~\ref{fig:p} 
one can appreciate that the vector components are spread up to that limit:
$u_{max}=14.3\ldots$. We suspect that the slope progressively decreases
as $N$ grows, approaching to $-1$ in the thermodynamic limit.
Another possibility is the existence of a prefactor in the power law
density. With an abuse of language: $\tilde P_s(u)\sim f(N)e^{-\alpha u}$,  
with $f(N)\sim N^{-1+\alpha}$. An extensive and
systematic numerical exploration might eventually clarify this issue.

\subsection{Numerical exploration of the Lyapunov exponent}

\begin{figure}
    \centering
    \includegraphics[width=0.9\linewidth]{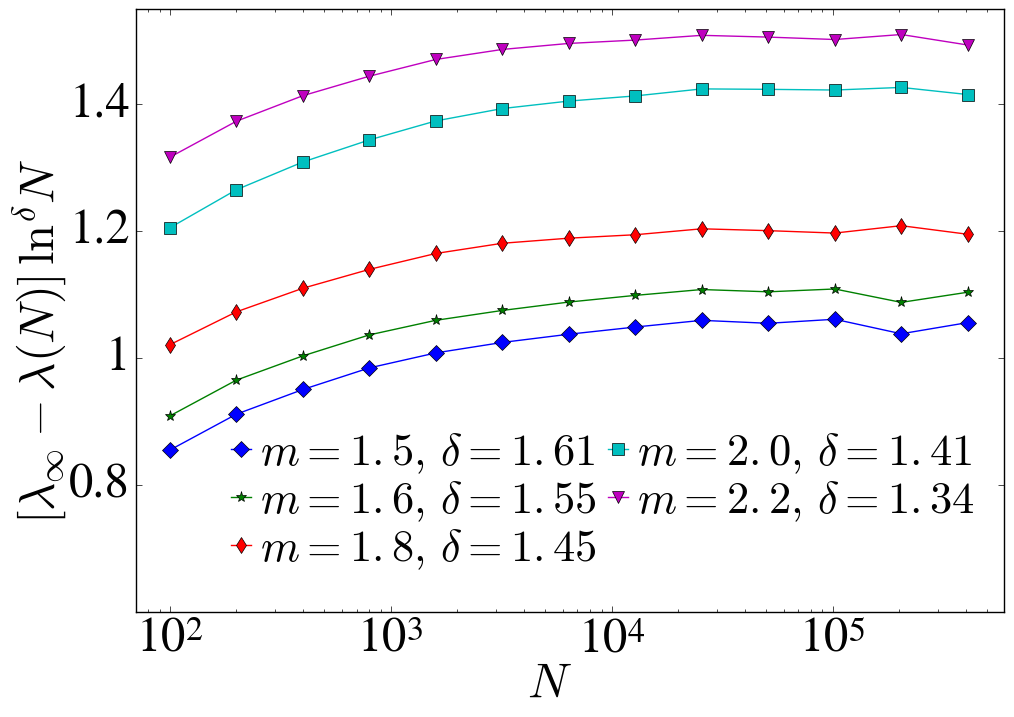}
    \caption{Rescaled LE difference $[\lambda_\infty-\lambda(N)] \ln^\delta N$ 
    for the {log-uniform RM model with $\epsilon=0.02$ and several
    values of the map parameter $m$.
    For each $m$ value, a different value of $\delta$ seems to be
    required to reach a plateau at large $N$.}}
    \label{fig:delta}
\end{figure}

One important consequence of the decay to
zero of the diffusion coefficient in regime III is that 
the constraint for $\lambda_\infty$ 
in Eq.~\eqref{cons} becomes the identity in Eq.~\eqref{linfty} 
(as in regimes I and II).
The knowledge of $\lambda_\infty=\ln\langle\mu\rangle$ persuaded us to explore 
the dependence of $\lambda(N)$ numerically.
Our guess is that the dependence must be logarithmic, as in the case of $d(N)$.
For comparison, { it is known that}, in one-dimensional spatio-temporal 
extensive chaos, $\lambda(L)$ and
$d(L)$ depend on the system size $L$ algebraically as
$L^{-1}$~\cite{pik98} and $L^{-1/2}$~\cite{kuptsov11,plp13}, respectively.
{ By analogy}, we attempted to accommodate our LE to
a generalized logarithmic scaling
\begin{equation}
\lambda_\infty -\lambda(N)\simeq \frac{c}{\ln^\delta N}
\label{lnd}
\end{equation}
Comparatively, our results in Fig.~\ref{fig:delta}
are much less robust than those for $d(N)$ in Fig.~\ref{fig:dif}.
As the system size increases the effective exponent $\delta$ grows
as well. We achieved systems sizes up to 
{ $N\approx 4\times10^{5}$} in Fig.~\ref{fig:delta}.
For each value of $m$, all of them in region III, 
a different value of $\delta$ flattens the curves
at large $N$ (notice the scaling with $\ln^\delta N$ in the $y$-axis). As previously mentioned, in all cases there is a systematic
increase of $\delta$ as $N$ grows. From our numerical results, we cannot discern
if $\delta$ takes a common asymptotic value or not.

\subsection{Rationale for a logarithmic law}

The lack of diffusion in the thermodynamic limit,
as well as a numerical check (not shown), indicates
that $s^t$ in Eq.~\eqref{ln_exact} fluctuates
around 1 with a decreasing amplitude as $N$ grows.
As before, this allows us expanding
the logarithm in Eq.~\eqref{init}, and subsequently 
deriving the expression
\begin{equation}
\lambda_\infty-\lambda(N) \propto \mathrm{var}(\tilde\mu) T_N  ,
\end{equation}
where $T_N$ is given by Eq.~\eqref{tn_av}, and $\tilde\mu$ was
defined in Sec.~\ref{sec:linfty}.
According to McLeish and O'Brien~\cite{mcleish82}, $T_N\sim(\ln N) ^{-1}$
if the tail index $\alpha$ equals unity.
This dependence would propagate up to $\lambda(N)$.
Note, however, that the result in~\cite{mcleish82} is fragile 
and breaks down under
a change in the drawing rule of the vector components: If they are
selected deterministically, then $T_N\sim (\ln N)^{-2}$.
In the case of the Lyapunov vector, 
components are indeed not completely independent as
the bound $v_j<N$ (if $\bar v=1$)  
immediately introduces certain correlations. 
Should we expect a convergence as $(\ln N)^{-\delta}$ for the LE 
with $1\le\delta\le2$? Is the numerical value of $\delta$ unique, or varies with the parameters? Unfortunately, we cannot give a proper answer to
these questions at this stage, and they are left as open problems.

\section{Discussion}
\label{sec:disc}

\subsection{Positive Multipliers: Regimes I, II, and III}
\label{sec:discb}

Our theoretical findings, built upon the random multiplier model
proposed by Takeuchi et al.~\cite{takeuchi11},
support the existence of three scaling regimes for turbulent
GCMs with positive multipliers. This result immediately
contradicts the very expectation of
a unique universal scaling law for the LE for this problem.
Regimes I and II exhibit power law behavior with different exponents.
For regime III, however, we could not determine 
the actual scaling properties, although
everything indicates it converges slower than a power law. Having the exact 
value of $\lambda_\infty$ allowed us to explore the conformity with 
a generalized logarithmic law, Eq.~\eqref{lnd}. But, clearly, 
more theoretical work is needed to better characterize GCMs in this regime.

\subsection{Implications for the general case: Positive and negative multipliers}

Our results do not apply to turbulent GCMs in which positive and negative
multipliers participate in the tangent dynamics. 
Nonetheless, our work immediately reveals that 
part of the analysis in~\cite{takeuchi11} is flawed. 
Let us enumerate, 
point-by-point, the key points leading us to
this important conclusion:
\begin{enumerate}
 \item The theoretical approximation
 in~\cite{takeuchi11} does not require any condition
 on the sign of the multipliers. As already explained, 
 the lack of sign-defined multipliers translates into Lyapunov-vector components
 with both signs.
 To cope with this, in \cite{takeuchi11}, the Hopf-Cole transformation simply 
 includes an absolute value: $u_i^t=\ln|v_i^t|$.
 Exclusively positive multipliers is 
 a best-case scenario, since the 
 Hopf-Cole transformation is invertible. If not, the problem
 is somehow brushed ``under the carpet''.
 
 \item For positive multipliers the density of the Lyapunov
 vector components rapidly decays to 0 as $v\to0$ (or as $u\to-\infty$),
 see Figs.~\ref{fig:Density_Pu} and \ref{fig:p}. This is consistent
 with the asymptotic behavior
 of the stationary solution of the Fokker-Planck 
 Eq.~\eqref{stationary}. After straightforward
 calculations we get: $\tilde P_s(u\to-\infty)\propto e^{-u^2/D}$ as $u\to-\infty$. 
 This abrupt decay is perceived as a lower wall in the density. Such a lower wall
 is invoked by Takeuchi et al.~\cite{takeuchi11}, see also Chap.~11 of~\cite{Pikovsky}, although in their
 reasoning finite-$N$ and infinite-$N$ perspectives are intermingled.
 In deep contrast, if positive and negative multipliers exist
 there is no lower wall. In the 
 empirical distribution shown in Fig.~\ref{fig:new} for a particular case (see caption), we 
 observe $\tilde P_s(u\to-\infty)\propto e^u$, implying~$P_s(v\to0)=\mathrm{const}$.
 This is not consistent with the Fokker-Planck equation. 
 The case of positive multipliers is, again, in better agreement with the
 ideas in \cite{takeuchi11}.
 
 \begin{figure}
	\centering
	\includegraphics[width=8.6cm]{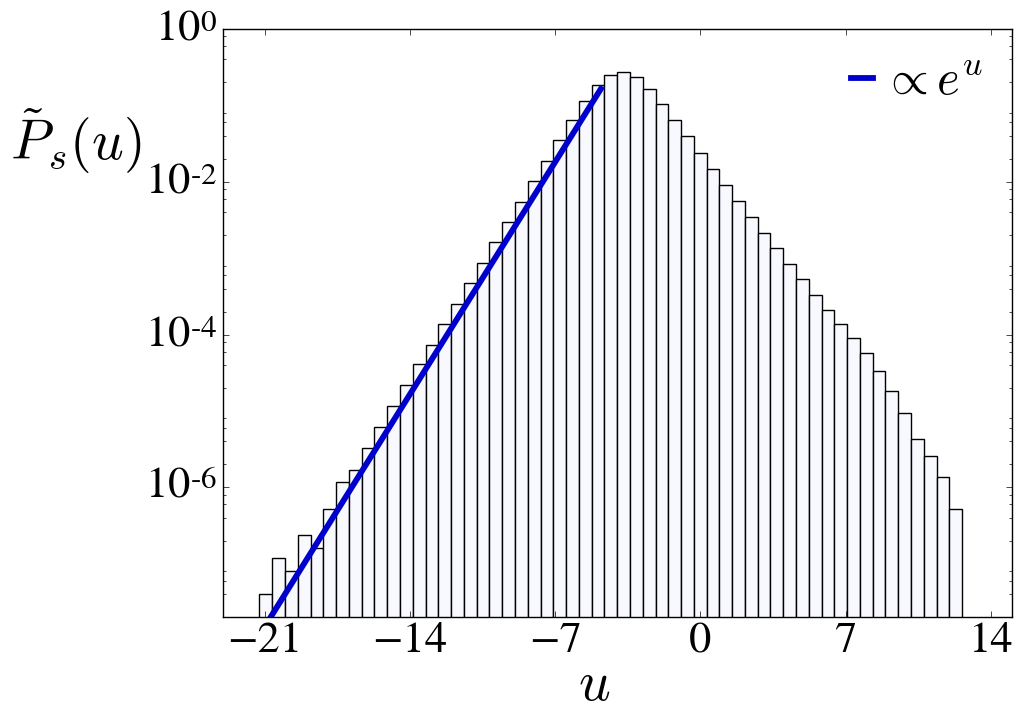}
	\caption{Probability density 
    $\tilde P_s(u)$ obtained form numerical simulations of the RM model
    with positive and negative multipliers distributed according to a
    bi-delta density: $\rho_{BD\pm}(\mu)=\frac1b\delta(\mu-b)
    +\frac{b-1}{b}\delta[\mu+b/(b-1)]$. This density corresponds
    to an isolated skewed-tent map.
    The parameters 
    are the same as in Fig.~\ref{fig:Density_Pu}: $\epsilon=0.02$, $b=3$.
    For small $u=\log|v|$, $\tilde P_s(u)\propto e^u$, 
    or equivalently $P_s(|v|\to0)=\mathrm{const}$.}
	\label{fig:new}
\end{figure}
 
 \item In spite of the apparent validity of the Fokker-Planck
 equation (at least, for positive multipliers), the
 predicted logarithmic scaling law in~\cite{takeuchi11}
 is in conflict with our results, which also relay on
 the Fokker-Planck equation. The origin of the
 discrepancy is elucidated next.

 \item  In the work by
Takeuchi et al.~the scaling law \eqref{ln} is 
exclusively derived from the Fokker-Planck equation.
This entails moving back and forth between finite and infinite $N$
cases. 
In particular, Takeuchi et al.~assumed that 
for finite $N$ the Fokker-Planck Eq.~\eqref{eq:fp} remains essentially
true replacing $\lambda_\infty$ by $\lambda(N)$. This assumption is crucial, but 
questionable, since the growth rate of the Lyapunov vector 
suffers fluctuations (i.e.,~diffusion), and there is not a co-moving reference frame
(for finite $N$).
In Ref.~\cite{takeuchi11}, changing $\lambda_\infty$ by $\lambda(N)$ modifies 
$\tilde P_s(u)$ as the decay rate $\alpha$ becomes $N$-dependent. 
The reasoning proceeds noticing that if $N$ is finite the maximal vector component
is about $u_{max}$ such that $\int_{u_{max}}^\infty \tilde P_s(u) du\sim N^{-1}$. 
For this scaling relation to hold true it is required that
\begin{equation}
\lambda(N)-\langle\ln \mu\rangle-\ln(1-\epsilon)\simeq\frac D2(1+c/\ln N) .
\label{tak}
\end{equation}
Unfortunately, this prediction is already erroneous at leading order. 
It implies $\lambda_\infty$ and $D$ are linked through
$\lambda_\infty-\langle\ln \mu\rangle-\ln(1-\epsilon)=\frac D2$,
and this yields $\alpha=1$ for $N\to\infty$, irrespective of the parameter values.
These implications are at odds with the numerical evidence, 
see e.g.~Fig.~\ref{fig:Density_Pu}. 
 
 \item  In our work, we have used the Fokker-Planck equation only to get
 the tail index $\alpha$ of $P_s(v)$ in the thermodynamic limit. The
 value of $\lambda_\infty$, and an analytic formula relating $\lambda(N)$
 with $\alpha$ and the multipliers' statistics
 are both derived here independently of the Fokker-Planck equation. This was
 possible thanks to the positiveness of the multipliers. 
  
 \item Our theory predicts the value of $\lambda_\infty$, which reduces the
 number of fitting parameters and makes us to be
 confident with the numerical tests of the theory.
 
 \item It might be argued that, apparently, the numerical results in \cite{takeuchi11} support
the logarithmic scaling law \eqref{ln}. There, however, 
$\lambda_\infty$ was a fitting parameter, what,
in our experience, 
introduces a substantial uncertainty and makes it impossible to really distinguish a logarithmic law from a 
power-law with small exponent $\gamma$. When $\lambda_\infty$ is theoretically known, as we have shown here, the power-law scaling can be established--- with an exponent that is model parameter dependent--- in regimes I and II, or even a more involved functional form in regime III.
\end{enumerate}

All in all, we conclude that the actual scaling law (or laws) 
for turbulent GCMs with multipliers 
adopting both signs remains an open problem.
For comparison, we note that in standard one-dimensional coupled-map 
lattices the sign of the
multipliers does not play a significant role.
In fact, for an asymptotically small coupling $\epsilon$,
the LE always varies as
$c/\ln\epsilon$ \cite{cecconi99}. 
If the sign of the multipliers 
changes or not only alters the scaling factor $c$. 
In addition, the convergence of the LE 
with the system size $N$ generically scales as $N^{-1}$. 
Remarkably, the theory is based on 
a mapping of the tangent 
dynamics to a simple stochastic partial-differential equation, in which all the Lyapunov vector components have the same sign~\cite{pik98,pazo08}.

Should we expect similar insensitivity of the 
LE with the multipliers sign in GCMs? 
This is an open problem that deserves a careful scrutiny. 
At this point our results constitute new evidence 
of the complex behavior of deceptively simple GCMs.

\begin{acknowledgments}
D.V.~acknowledges support by Agencia Estatal de Investigaci\'on 
and Fondo Social Europeo (Spain) under the doctoral fellowship No.~BES-2017-081808 
of the FPI Programme.
We acknowledge support by Agencia Estatal de Investigaci\'on and 
Fondo Europeo de Desarrollo Regional under Project 
No.~FIS2016-74957-P (AEI/FEDER, EU).
\end{acknowledgments}

\setcounter{equation}{0} 
\renewcommand{\theequation}{A\arabic{equation}}

\section*{Appendix: Derivation of Eq.~(30) and numerical tests}

\begin{figure}[h!]
	\centering
	\includegraphics[width=8.6cm]{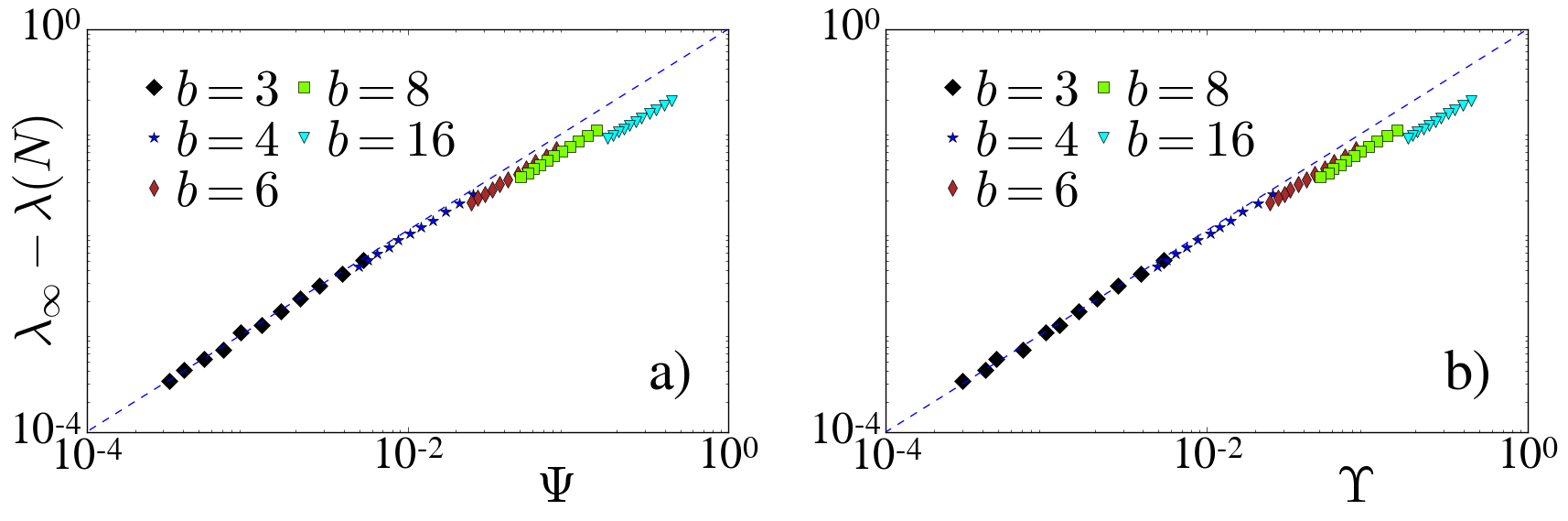}
	\caption{Numerical tests of 
		Eq.~\eqref{average} (a) and  Eq.~\eqref{final} 
		(b) for the {bi-delta RM model 
		with $\epsilon=0.02$ and $b=\{3, 4, 6, 8,16\}$.} 
		(a) The label $\Psi$ in the $x$-axis stands for the right-hand side of 
		Eq.~\eqref{average}. The data progressively deviate from the bisectrix (dashed line) as $b$ increases. For each value of $b$ the left most point corresponds 
		to the largest size $N=102400$ 
		(b) The same as (a), but now $ \Upsilon$ in the $x$-axis
		denotes the right-hand side
		of Eq.~\eqref{final}.}
	\label{fig:tests}
\end{figure}

For simplicity, we start approximating $\bar \mu$ by $\langle\mu\rangle$
in Eq.~\eqref{init}, and re-write it
in terms of the deviations from the expected muliplier value 
$\delta\mu_j^t=\mu_j^t-\langle\mu\rangle$:
\begin{equation}
 \lambda(N) \simeq\lambda_\infty  +
 \left<\ln\left[1+ \frac{1-\epsilon}{\langle\mu\rangle } 
 \frac{\sum_{j=1}^{N} \delta\mu_{j}^t v_{j}^t}{\sum_{j=1}^{N}  v_{j}^t}\right]\right> 
\label{exact}
 \end{equation}
{If $N$ is large we can Taylor expand the logarithm: $\ln(1+x)=x-x^2/2+\cdots$,
where} the first order of the expansion is zero, and 
we truncate at second order:
\begin{equation}
 \lambda_\infty-\lambda(N) \simeq \frac{(1-\epsilon)^2}{2 \langle\mu\rangle^2}
 \left<\left( \frac{\sum_{j=1}^{N} \delta\mu_{j}^t v_{j}^t}{\sum_{j=1}^{N} v_{j}^t}   \right)^2\right> 
\label{ho}
 \end{equation}
The numerator of the average in the right-hand side can be further simplified. 
Given that the $\delta\mu_j^t$ are
completely uncorrelated and have zero mean, we can expand the square and cancel out all 
(covariance-like) cross-products. Doing so we get this expression:
\begin{equation}
\lambda_\infty- \lambda(N) \simeq \frac{(1-\epsilon)^2}{2 \langle\mu\rangle^2}  
\left<  \frac{\sum_{j=1}^{N} { (\delta\mu_{j}^t)^2} (v_{j}^t)^2}{\left(\sum_{j=1}^{N} v_{j}^t   \right)^2}\right>
\label{average}
\end{equation}
The average in the right-hand side, denoted $\Psi$ for short,  is far from trivial. As a first check, we represent in Fig.~\ref{fig:tests}(a)
the two sides of this equation with data obtained from numerical simulations of the RM model
with the bi-delta density. Several values of parameter $b$ were selected as well
as a common coupling constant $\epsilon=0.02$. For the larger $b$ values 
we may appreciate deviations
from the bisectrix, which can only be attributed to the slow convergence
of the Taylor expansion of the logarithm. Indeed, our numerical tests of
Eq.~\eqref{exact}
yield a quasi perfect agreement between theory and data for all $b$ values. 
This evidences the risk of relying on numerical simulations alone,
given that the asymptotic regime shows up at prohibitively large system sizes 
for some parameter values.
In the particular case of the bi-delta density, it is already difficult to observe 
the asymptotic decay for $b$ above $4$.

If the multipliers do not exhibit large fluctuations
we can approximate ${ (\delta\mu_{j}^t)^2}$ by the variance of $\mu$,
and obtain Eq.~\eqref{final}
In Fig.~\ref{fig:tests}(b)
we test Eq.~\eqref{final} setting $\epsilon=0.02$ and 
the bi-delta density \eqref{bidelta} for several values of $b$.
The results are comparable to those in Fig.~\ref{fig:tests}(a), 
evidencing that putting the variance of $\mu$ out of the average
does not deteriorate the accuracy of the approximation.


%

\end{document}